\documentclass[preprint,showpacs,12pt,floatfix,aps]{revtex4}

\usepackage{graphicx}
\usepackage{latexsym}
\usepackage{amsmath}
\input psfig.sty

\newcommand{\bq}{\begin{equation}}
\newcommand{\eq}{\end{equation}}
\newcommand{\bqn}{\begin{eqnarray}}
\newcommand{\eqn}{\end{eqnarray}}
\newcommand{\nb}{\nonumber}
\newcommand{\lb}{\label}
\newcommand{\rr}{\bf r}
\begin{document}
\title{Charged Gravastar in a Dark Energy Universe}
\author{C.F.C. Brandt $^{1}$}
\email{fredcharret@yahoo.com.br}
\author{R. Chan $^{2}$}
\email{chan@on.br}
\author{M.F.A. da Silva $^{1}$}
\email{mfasnic@gmail.com}
\author{P. Rocha $^{13}$}
\email{pedrosennarocha@gmail.com}
\affiliation{$^{1}$ Departamento de F\'{\i}sica Te\'orica,
Instituto de F\'{\i}sica, Universidade do Estado do Rio de Janeiro,
Rua S\~ao Francisco Xavier 524, Maracan\~a,
CEP 20550-900, Rio de Janeiro - RJ, Brasil\\
$^{2}$ Coordena\c{c}\~ao de Astronomia e Astrof\'{\i}sica, Observat\'orio
Nacional, Rua General Jos\'e Cristino, 77, S\~ao Crist\'ov\~ao, CEP 20921-400,
Rio de Janeiro, RJ, Brazil\\
$^{3}$ IST - Instituto Superior de Tecnologia de Paracambi, FAETEC, 
Rua Sebasti\~ao de Lacerda
s/n, Bairro da F\'abrica, Paracambi, 26600-000, RJ, Brazil}
 
\date{\today}

\begin{abstract}
Here we constructed a charged gravastar model formed by an interior
de Sitter spacetime, a charged dynamical infinitely thin  shell with
an equation of state and an exterior  de Sitter-Reissner-Nordstr\"om
spacetime. We find that the presence of the charge is crucial to the
stability of these structures. It can as much favor the stability of a
bounded excursion gravastar, and still converting it in a stable gravastar,
as make disappear a stable gravastar, depending on the range of the charge
considered. There is also formation of black holes and, above certain
values, the presence of the charge allows the formation of naked
singularity . This is an important example in which a naked singularity
emerges as a consequence of unstabilities of a gravastar model, which
reinforces that gravastar is not an alternative model to black hole.
\end{abstract}


\maketitle

\section{Introduction}

Although we have strong theoretical and experimental evidences in favor of
the existence of black holes,  lots of paradoxical problems about them do
exist\cite{Wald01}.  Besides, it was shown recently that observational
data can give strong arguments in the existence of event horizons,
but we can not prove it directly \cite{AKL02}.

We also have the fact that the picture of gravitational collapse provided by Einstein's General
Relativity cannot be completely correct since in the final stages of collapse quantum effects
must be taken into account at high curvature values, or short distances, compared with the Planck length
scale, as pointed out by Chapline \cite{chapline} and others researchers.

These facts frequently motivated authors to try to find new alternatives for
the final state of a collapsing star without horizons. Among these models we
can mention Bose superfluid\cite{CHLS03}, dark stars \cite{Vach07} and holostars 
\cite{holostar1,holostar2,holostar3,holostar4,holostar5,holostar6}.

There are many others additional models proposed as black holes mimickers, but among the
alternative models to these compact objects, the gravitational vacuum stars (gravastars) \cite{MM01}
received special attention, partially due to the tight connection  between the cosmological constant and our accelerated
expanding universe, although it is very difficult to distinguish these objects from black holes.

In the original model of Mazur and Mottola (MM) \cite{MM01}, gravastars consist of five layers. To study the dynamical stability of such compact object, Visser and Wiltshire (VW) \cite{VW04} argued that such five-layer models can be simplified to three-layer ones. They also pointed out that there are two different types of stable gravastars which are stable gravastars and "bounded excursion" gravastars.
The first one represents a stable structure already formed, while the second one is a system with a shell which oscillates around a equilibrium position which can loose energy and stabilizing at the end.

Recently we have studied the stability of some three layer gravastar models\cite{JCAP,JCAP1,JCAP2,JCAP3,JCAP4,GRG}.
The first model \cite{JCAP} consisted of an internal de Sitter spacetime, a
dynamical infinitely thin  shell of stiff fluid, and an external Schwarzschild spacetime, as proposed by VW
\cite{VW04}.
We have shown explicitly that the final output can be a black hole, a "bounded excursion" stable gravastar, a Minkowski, or a de Sitter
spacetime, depending on the total mass $m$ of the system,  the cosmological constant
$\Lambda$ and the initial position $R_{0}$ of the dynamical shell. We have concluded that 
although it does exist a region of the space of the initial parameters where it is always formed
stable gravastars, it still exists a large region of this space where
we can find black hole formation. In the sequence, considering after an
equation of state $p = (1-\gamma)\sigma$ for the shell \cite{JCAP1}, instead
of only using a stiff fluid ($\gamma=0$), we concluded that gravastar really
is not an alternative model to black hole.

We also have generalized the former one considering an interior constituted by an anisotropic dark energy fluid \cite{JCAP2,GRG}.
We have again confirmed the previous results, i.e., that both gravastars and
black holes can be formed, depending on the initial parameters. 
It is remarkable that for this case we have an interior fulfilled by a
physical matter,
instead of a de Sitter vacuum.  Thus, it is similar to dark energy star models.

Recently, Carter \cite{Carter} studied spherically symmetric gravastar
solutions which possess an
(anti) de Sitter interior and a (anti) de Sitter-Schwarzschild or
Reissner-Nordstr\"om exterior, separately. He followed the same approach that
Visser and Wiltshire took in their work \cite{VW04} assuming a potential $V(R)$ and
then founding the equation of state of the shell. He found a wide range
of parameters which allows stable gravastar solutions, and presented the
different qualitative behaviors of the equation of state for these
parameters, for both cases, that are, a (anti) de Sitter-Schwarzschild or
Reissner-Nordstr\"om exterior.

Differently from Carter's work \cite{Carter}, we consider here a different approach, as in the
previous works. In a first step, we generalized our second work on gravastars \cite{JCAP1}, introducing an external de
Sitter-Schwarzschild spacetime \cite{JCAP3}. The aim was to study how the cosmological
constant affects the gravastar formation, and we found that the exterior
cosmological constant imposes a limit on the gravastar formation since the
dark energy density inside the gravastar has to be greater than the surrounding
spacetime. Now we are interested in the influence of the charge, combined with
the influence of an exterior cosmological constant, considering
a de Sitter-Reissner-Nordstr\"om exterior spacetime. For this configuration
we showed that the presence of the charge can changes considerably the
stability conditions of these structures. It can as much favor the stability
of a bounded excursion gravastar, converting it in a stable gravastar, as make
disappear a stable gravastar or even to allow a naked singularity formation.

In a previous work \cite{JCAP4} we have already considered the exterior of the
gravastar is a
Reissner-Nordstr\"om spacetime, but with zero total mass and, depending on the
parameter $\omega=1-\gamma$ of the equation of state of the shell, and the
charge, a gravastar structure can be formed. We have found that the presence
of the charge contributes to the stability of the gravastar, if the charge is
greater than a critical value.

The paper is  organized as follows: In Sec. II we present the metrics of the
interior and exterior spacetimes, with theirs extrinsic curvatures,
the equation of motion of the shell and the potential of the system.
In Sec. III we analyze the influence of the presence of the charge in the
gravastar model confirming the existence of naked singularity formation and
we investigate the formation of
gravastar from numerical analysis of the general potential.
Finally, in Sec. IV we present our conclusions.

\section{Formation of Gravastars in a de Sitter - Reissner - Nordstr\"om spacetime}

This gravastar model is described by an interior spacetime with a cosmological constant $\Lambda_i$, given by  the de Sitter metric, 
\bq
ds^2_{i}=-\left(1-\frac{r^2}{L_i^2}\right) dt^2 + \left(1-\frac{r^2}{L_i^2}\right)^{-1} dr^2 + r^2 d\Omega^2,
\lb{ds2-}
\eq
where, $L_i=\sqrt{3/\Lambda_i}$ and 
$d\Omega^2 = d\theta^2 + \sin^2(\theta)d\phi^2$.

Differently from the previous models, here we consider a charged shell and a cosmological constant $\Lambda_e$ generating a vacuum  exterior spacetime described by the de  
Sitter-Schwarzschild-Reissner-Nordstr\"om metric, 
\bq
ds^2_{e}= -\left(1-\frac{2m}{r}-\frac{r^2}{L_e^2}+\frac{q^2}{r^2}\right) dv^2 + \left(1-\frac{2m}{r}-\frac{r^2}{L_e^2}+\frac{q^2}{r^2}\right)^{-1} d{\rr}^2 + {\rr}^2 d\Omega^2,
\lb{ds2+}
\eq
where $L_e=\sqrt{3/\Lambda_e}$. 

The thin shell is characterized by the hypersurface $r=r_{\Sigma}$  and given by the metric
\bq
ds^2_{\Sigma}= -d\tau^2 + R^2(\tau) d\Omega^2,
\lb{ds2Sigma}
\eq
where $\tau$ is the proper time.

In order to find the mass of the shell, and then its potential, it is necessary to consider the junction conditions.

The continuity of the first fundamental form impose that $ds^2_{i} = ds^2_{e} = ds^2_{\Sigma}$, then $r_{\Sigma}={\rr}_{\Sigma}=R$
and
\bqn
\dot t^2&=&\left[ f_1 - f_1^{-1} \left( \frac{\dot R}{\dot t} \right)^2 \right]^{-1}= \nb \\
& & \left[ 1 - \left(\frac{R}{L_i}\right)^2 \right]^{-2} \left[1 - \left(\frac{R}{L_i}\right)^2
 + \dot{R}^2 \right] ,
\lb{dott2}
\eqn
and
\bqn
\dot v^2&=&\left[ f - f^{-1} \left( \frac{\dot R}{\dot v} \right)^2 \right]^{-1}= \nb \\
& & \frac{1}{f^2} \left[ \dot{R}^2 + 1 -2 \frac{m}{R} - \left(\frac{R}{L_e}\right)^2
+ \left(\frac{q}{R}\right)^2 \right]  ,
\lb{dotv2}
\eqn
where the dot represents the differentiation with respect to $\tau$.

Thus, the interior and exterior normal vector are given by
\bq
n^{i}_{\alpha} = (-\dot R, \dot t, 0 , 0 ),
\lb{nalpha-}
\eq
and
\bq
n^{e}_{\alpha} = (-\dot R, \dot v, 0 , 0 ).
\lb{nalpha+}
\eq

The interior and exterior extrinsic curvature are given by
\bqn
K^{i}_{\tau\tau}&=&-\left[(3 L_i^4 \dot R^2 - L_i^4 \dot t^2 + 2 L_i^2 R^2 \dot t^2 -
R^4 \dot t^2) R \dot t \right.\nb \\
& &\left. -(L_i+R) (L_i-R) (\dot R \ddot t - \ddot R \dot t) L_i^4\right] 
\left[(L_i+R) (L_i-R) L_i^4\right]^{-1}
\lb{Ktautau-}
\eqn
\bq
K^{i}_{\theta\theta}=\dot t (L_i+R) (L_i-R) L_i^{-2} R
\lb{Kthetatheta-}
\eq
\bq
K^{i}_{\phi\phi}=K^{i}_{\theta\theta}\sin^2(\theta),
\lb{Kphiphi-}
\eq
\bqn
K^{e}_{\tau\tau}&=& - \dot{v} \left[ \left(q^2 + R^2 - 2 m R \right) L_e^2 - R^4\right]^{-1}
\left[4 L_e^4 m^2 R^2 \dot{v}^2 - 4 L_e^4 m q^2 R \dot{v}^2\right. \nb \\
& &  - 4 L_e^4 m R^3 \dot{v}^2 
+ L_e^4 q^4 \dot{v} +  2 L_e^4 q^2  r^2 \dot{v}^2 - 3 L_e^4 R^4 \dot{R}^2+ L_e^4 R^4 \dot{v}^2 
\nb \\
& & \left.+ 4 L_e^2 m R^5 \dot{v} -2 L_e^2 q^2 R^4 \dot{v}^2 - 2 L_e^2 R^6 \dot{v}^2+R^8 \dot{v}^2\right] \nb\\
& & \left(L_e^2 m R - L_e^2 q^2 - R^4\right) L_e^{-4} R^{-5} 
+ \dot{R}\ddot{v} - \ddot{R} \dot{v}
\lb{Ktautau+}
\eqn
\bq
K^{e}_{\theta\theta}=\dot v \left[\left(q^2 +R^2 - 2 m R \right) L_e^2 - R^4\right] L_e^{-2} R^{-1}
\lb{Kthetatheta+}
\eq
\bq
K^{e}_{\phi\phi}=K^{e}_{\theta\theta}\sin^2(\theta).
\lb{Kphiphi+}
\eq

Following Lake \cite{Lake}, we have
\bq
[K_{\theta\theta}]= K^{e}_{\theta\theta}-K^{i}_{\theta\theta} = - M,
\lb{M}
\eq
where $M$ is the mass of the shell. Thus
\bq
M = \dot{t} (L_i + R)(L_i  - R) \frac{R}{L_i^2} - \dot{v} \left[ (q^2 + R^2 - 2m R) L_e^2
- R^4 \right] \frac{1}{R L_e^2}.
\lb{M1}
\eq

Then, substituting equations (\ref{dott2}) and (\ref{dotv2}) into (\ref{M1}) 
we get
\bq
M - R \left[ 1 -\left(\frac{R}{L_i}\right)^2  + \dot R^2 \right]^{1/2}
+ 
R  \left[1-\frac{2m}{R} - \left(\frac{R}{L_e}\right)^2 + 
\dot R^2  + \left(\frac{q}{R}\right)^2\right]^{1/2}  
=0.
\lb{M2}
\eq

Solving the equation (\ref{M2}) for $\dot R^2$ we obtain the potential 
$V(R,m,q,L_i,L_e)$ \cite{JCAP1}.
In order to keep the ideas of our work \cite{JCAP1} as much as possible, we 
consider the thin shell as consisting
of a fluid with a equation of state, $p = (1-\gamma)\sigma$, where 
$\sigma$ and $p$ denote, 
respectively, the surface energy density and pressure of the shell and 
$\gamma$ is a constant. 
The equation of motion of the shell is given by \cite{Lake}
\bq
\dot M + 8\pi R \dot R \vartheta = 4 \pi R^2 [T_{\alpha\beta}u^{\alpha}n^{\beta}]=
4 \pi R^2 \left(T^e_{\alpha\beta}u_e^{\alpha}n_e^{\beta}-T^i_{\alpha\beta}u_i^{\alpha}n_i^{\beta} \right),
\lb{dotM}
\eq
where $u^{\alpha}$ is the four-velocity.  Since the interior region is
constituted by a fluid with cosmological constant and the exterior corresponds 
to a charged spacetime characterized by the Reissner-Nordstr\"om with exterior
cosmological constant, we get
\bq
\dot M + 8\pi R \dot R (1-\gamma)\sigma = 0,
\lb{dotM1}
\eq
since \cite{Adler}
\bqn
T_{\alpha\beta}^e & = & F_{\alpha\lambda}F^{\lambda}_{\beta}+g_{\alpha\beta}\frac{1}{4}F_{\lambda\nu}F^{\lambda\nu}+\Lambda_e g_{\alpha\beta}\nb \\
& = & \frac{q^2}{2\rr^4}\left( f\delta^{v}_{\alpha}\delta^{v}_{\beta}-
f^{-1}\delta^{\rr}_{\alpha}\delta^{\rr}_{\beta}+
\rr^2\delta^{\theta}_{\alpha}\delta^{\theta}_{\beta}+
\rr^2 \sin^2\theta \delta^{\phi}_{\alpha}\delta^{\phi}_{\beta} \right)+\Lambda_e g_{\alpha\beta},
\eqn
where $F_{\alpha\lambda}$ is the Maxwell tensor.

Since $\sigma = M/(4\pi R^2)$ we can solve equation (\ref{dotM1}) giving
\bq
M=k R^{2(\gamma-1)},
\lb{Mk}
\eq
where $k$ is an integration constant.

Substituting equation (\ref{Mk}) into $V(R,m,q,L_i,L_e)$ we 
obtain the general expression for the potential, 
\bqn
\lb{VR}
V(R,m,q,L_i,L_e,k,\gamma) & = & \frac{1}{2} - \frac{1}{4}{\frac {{r}^{2}}{{{\it L_i}}^{2}}}
  - \frac{1}{8}{\frac {{R}^{10}}{{k}^{2} {R}^{4\gamma} {\it L_i}^{4}}} + \frac{1}{4}{\frac {{q}^{2}}{
{R}^{2}}} - \frac{1}{2}{\frac {{\it m}}{R}} 
             +   \frac{1}{2}{\frac {{R}^{3}{\it m}\,{q}
^{2}}{{k}^{2} {R}^{4\gamma} }}  \nb\\ 
  & - & \frac{1}{4}{\frac {{R}^{2}}{L_e^{2}}} - \frac{1}{8}{\frac {{R}^{10}}{{k}^{2} {R}^
{4\gamma} L_e^{4}}} + \frac{1}{4}{\frac {{R}^{10}}{{k
}^{2} {R}^{4\gamma} L_e^{4}}} \nb\\
 & - & \frac{1}{8}{
\frac {{k}^{2} {R}^{4\gamma} }{{R}^{6}}} - \frac{1}{4}{\frac {{R}
^{6}{q}^{2}}{{k}^{2} {R}^{4\gamma} {{\it L_i}}^{2}}} 
  - \frac{1}{8}
{\frac {{R}^{2}{q}^{4}}{{k}^{2} {R}^{4\gamma} }} + \frac{1}{4}{
\frac {{R}^{6}{q}^{2}}{{k}^{2} {R}^{4\gamma} L_e^{2}}} \nb\\ 
 & + & \frac{1}{2}{\frac {{R}^{7}{\it m}}{{k}^{2} {R}^{4\gamma}
 {{\it L_i}}^{2}}} - \frac{1}{2}{\frac {{R}^{7}{\it m}}{{k}^{2}
 {R}^{4\gamma} L_e^{2}}} - \frac{1}{2}{\frac {
{R}^{4}{{\it m}}^{2}}{{k}^{2} {R}^{4\gamma} }}
\eqn

Redefining the Schwarzschild mass $m$, the charge $q$, the cosmological 
constants $L_i$ and $L_e$ and the radius $R$ as
\bq
m \rightarrow mk^{-\frac{1}{2\gamma-3}},
\eq
\bq
q \rightarrow qk^{\frac{2}{2\gamma-3}},
\eq
\bq
L_i \rightarrow L_i k^{\frac{2}{2\gamma-3}},
\eq
\bq
L_e \rightarrow L_e k^{\frac{2}{2\gamma-3}},
\eq
\bq
R \rightarrow Rk^{-\frac{1}{2\gamma-3}},
\eq
we get the potential
\bqn
\lb{VRa}
V(R,m,q,L_i,L_e,\gamma) & = & \frac{1}{2} - \frac{1}{4}{\frac {{R}^{2}}{{{\it L_i}}^{2}}}
  - \frac{1}{8}{\frac {{R}^{10}}{ {R}^{4\gamma} {\it L_i}^{4}}} + \frac{1}{4}{\frac {{q}^{2}}{
{R}^{2}}} - \frac{1}{2}{\frac {{\it m}}{R}} 
             +   \frac{1}{2}{\frac {{R}^{3}{\it m}\,{q}
^{2}}{ {R}^{4\gamma} }}  \nb\\ 
  & - & \frac{1}{4}{\frac {{R}^{2}}{L_e^{2}}} - \frac{1}{8}{\frac {{R}^{10}}{ {R}^
{4\gamma} L_e^{4}}} + \frac{1}{4}{\frac {{R}^{10}}{
 {R}^{4\gamma} L_e^{4}}} \nb\\
 & - & \frac{1}{8}{
\frac { {R}^{4\gamma} }{{R}^{6}}} - \frac{1}{4}{\frac {{R}
^{6}{q}^{2}}{ {R}^{4\gamma} {{\it L_i}}^{2}}} 
  - \frac{1}{8}
{\frac {{R}^{2}{q}^{4}}{ {R}^{4\gamma} }} + \frac{1}{4}{
\frac {{R}^{6}{q}^{2}}{ {R}^{4\gamma} L_e^{2}}} \nb\\ 
 & + & \frac{1}{2}{\frac {{R}^{7}{\it m}}{ {R}^{4\gamma}
 {{\it L_i}}^{2}}} - \frac{1}{2}{\frac {{R}^{7}{\it m}}{
 {R}^{4\gamma} L_e^{2}}} - \frac{1}{2}{\frac {
{R}^{4}{{\it m}}^{2}}{ {R}^{4\gamma} }}
\eqn

Therefore, for any given constants $m$, $q$, $L_i$, $L_e$ and $\gamma$, 
equations (\ref{VR}) or (\ref{VRa}) uniquely determines the collapse
of the  shell.
Observe that the exponents of the charge, as well as those of $L_i$ and $L_e$,  are always even implying that its sign
is irrelevant.

The gravastar model constructed here shows 4 different horizons, which are:

\bq
R_{dsec}=-Z_1+Z_2+Z_3,
\lb{Rdsec}
\eq
\bq
R_{rnoah}=Z_1-Z_2+Z_3,
\lb{Rrnoah}
\eq
\bq
R_{rniah}=Z_1+Z_2-Z_3,
\lb{Rrniah}
\eq
\bq
R_{dsic}=L_i,
\lb{Rdsic}
\eq
where
\bq
Z_1=\frac{1}{\sqrt{2\Lambda_e}}\sqrt{1-\sqrt{1-4\Lambda_e q^2}\cos(\psi/3-\pi/3)},
\eq
\bq
Z_2=\frac{1}{\sqrt{2\Lambda_e}}\sqrt{1-\sqrt{1-4\Lambda_e q^2}\cos(\psi/3+\pi/3)},
\eq
\bq
Z_3=\frac{1}{\sqrt{2\Lambda_e}}\sqrt{1+\sqrt{1-4\Lambda_e q^2}\cos(\psi/3)},
\eq
and
\bq
\psi=\arccos \left(-y \right),
\eq
where
\bq
y=\frac{1-18 m^2 \Lambda_e+12 q^2 \Lambda_e}{(1-4 q^2 \Lambda_e)^{3/2}},
\eq
where $\Lambda_e=3/L_e^2$, $R_{dsec}$ is the exterior cosmological horizon, $R_{rnoah}$ is the
outer apparent horizon, $R_{rniah}$ is the inner apparent horizon, all of them for the de-Sitter-Reissner-Nordstr\"om exterior spacetime \cite{Koberlein} and $R_{dsic}$ is the
cosmological horizon for the interior de Sitter spacetime.
Note that if $y > 1$ this gives us an imaginary angle $\psi$.
In this case, the horizons $R_{dsec}$, $R_{rnoah}$ and $R_{rniah}$ are imaginary.  Since any spacetime is defined only
for real and non-negative radii, horizons obtained from equations (\ref{Rdsec})-(\ref{Rdsic})
can not be negative or imaginary. When this occurs it means that the spacetime is
horizon-free.

To  guarantee that initially the spacetime does not have any kind of horizons,
cosmological or event, in general,  we must restrict $R_{0}$ to the ranges simultaneously,

$R_0 > R_{rnoah}$,

and

$R_0 < R_{dsic}$, if $L_i < L_e$, 

\noindent where $R_0$ is the initial collapse radius.

In the particular case of $q=0$, the exterior horizons are given by 
\cite{Shankaranarayanan}
\bq
R_{bh}= \frac{2m}{\sqrt{3 z}} \cos \left( \frac{\pi+\Psi}{3} \right),
\label{rbh}
\eq
\bq
R_c= \frac{2m}{\sqrt{3 z}} \cos \left( \frac{\pi-\Psi}{3} \right),
\label{rc}
\eq
where $z=(m/L_e)^2$, $\Psi= \arccos \left( 3\sqrt{3 z} \right)$, $R_{bh}$ denotes the black
hole horizon and $R_c$ denotes the cosmological horizon.
Note that if $z > 1/27$
the quantity $3\sqrt{3 z}$ is greater than 1, giving an imaginary angle $\Psi$.
Thus, the
horizons $R_{bh}$ and $R_c$ are imaginary and the spacetime becomes free of
horizons.

Since $\sigma\geq 0$, in order to avoid dark energy fluids, we must to have 
$\sigma+2P\geq 0$ for the shell
and assuming that $p=(1-\gamma)\sigma$ we must have            
$\gamma \le 1.5$. On the other hand, in order
to satisfy the condition $\sigma+p\ge 0$, we get that $\gamma \le 2$.
The dominant energy condition is only satisfied for $0 \le \gamma \le 2$.
Although the phantom energy is usually considered as a kind of dark energy,
in this paper we will use the expression dark energy for the case where the
condition $\sigma + p \ge 0$ is satisfied and phantom energy otherwise.
Hereinafter, we will use only some particular values of the parameter
$\gamma$ which are analyzed in this work. See Table \ref{table0}.

\begin{table}
\caption{\label{table0}This table summarizes the matter classification
based on the energy conditions of the shell, in terms of the parameter $\gamma$.}
\begin{ruledtabular}
\begin{tabular}{ccccc}
Matter & Condition 1 & Condition 2  & $\gamma$ & $\gamma$ of this work\\
\hline
Standard Energy           & $\sigma+2p\ge 0$ & $\sigma+p\ge 0$ & $\gamma \le 1.5$  & 0, -1\\
Dark Energy               & $\sigma+2p\le 0$ & $\sigma+p\ge 0$ & $1.5 \le \gamma \le 2$ & 1.7 \\
Phantom Energy  & $\sigma+2p\le 0$ & $\sigma+p\le 0$ & $\gamma \ge 2$ & 3\\
\end{tabular}
\end{ruledtabular}
\end{table}

\section{General Case}

In the following, have done a graphical analysis of  several special cases.
The influence of the cosmological constant was deeply discussed in  a previous work \cite{JCAP3}, where $L_i\neq\infty$ and
$L_e\neq\infty$, but $q=0$, we showed that there is a limit on $L_e$ in
order to form a gravastar, i.e.,
the formation of gravastars depends on the value of $L_e$
($L_e > {L_e}^{min}$, with ${L_e}^{min}\geq L_i$) in a such way that,
instead of what occurs for $L_i$, as smaller is $L_e$ as bigger is
the tendency to the collapse.
Now, our main aim is to study the role of the charge in the dynamic of the gravastar. 
Our strategy is to start with values of $m$, $\gamma$,
$L_i$ and $L_e$, for which we had a bounded excursion or a stable
gravastar for chargeless configurations and, then, gradually introduce and increase the value of the charge. Thus, we
can investigate if there is a range for the charge favoring the formation any kind  of structure.

In order to analyze the effect of the charge we have started from the cases with q=0, 
considered in the previous work \cite{JCAP3}, and we have used the critical mass, when 
a stable gravastar was formed, changing the value of the charge. 
Recalling that the critical mass is defined as the mass for which $V(R)=0$ and $dV(R)/dR=0$,
for a fixed value of the charge. 
The results are shown 
in the Tables II-VI, and the respective potential are in the Figure 1 and Figures 4-7. 
In particular, in the Figure 1, we have lost the stable configuration when
we increase the value of the charge.  In order to check if it is possible to 
have a charged stable gravastar we plotted the Figure \ref{fig1b}.  In this figure we
have fixed the value of the mass and we searched for the critical charge.
The critical charge is defined as the charge for which $V(R)=0$ and $dV(R)/dR=0$, for a
fixed value of the mass. It is remarkable the crucial role of the charge. 
Note that there is an approximated interval $1 < q < 1.17$ in which we can always find 
bounded 
excursion or stable gravastar, where the value $q=1.17$ corresponds to the stable gravastar.
For the values outside of this interval, we always have the collapse of the shell.
In the Figure \ref{fig1a} we show that a similar interval for the mass can be also found for a
fixed value of charge ($0.8795 < m < 1$).
Thus we can have stable or bounded excursion gravastar 
depending on the combination of charge and mass.

We call attention that in all the cases studied here, the formation of the apparent horizon can 
be avoided increasing the value of the charge, indicating that the shell can collapse to form naked singularity.  

\begin{figure}
\vspace{.2in}
\centerline{\psfig{figure=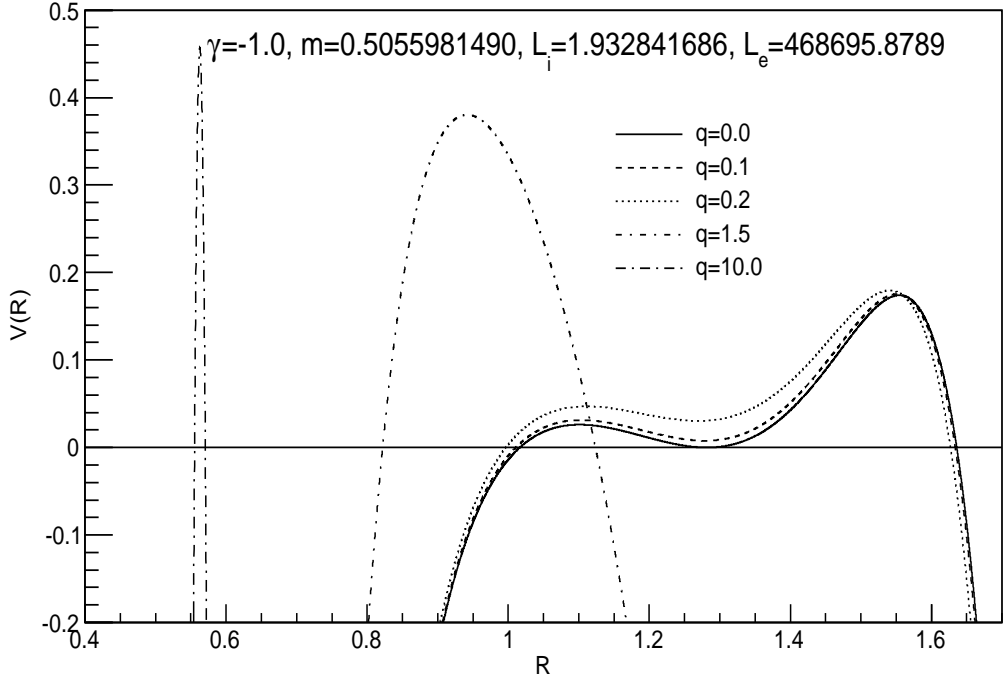,width=6.0truein,height=4.0truein}
}\caption{The potential $V(R)$ for $\gamma=-1$, $m=0.5055981490$, 
$L_i=1.932841686$ and $L_e=468695.8789$. The horizons are given in Table \ref{table1}. 
The growth of the charge eliminates the stable structure. In addition, note that for $q=1.5$, the shell collapses to a black hole, while for $q=10.0$, we have the collapse of the shell in a naked singularity.}
\label{fig1}
\end{figure}

\begin{figure}
\vspace{.2in}
\centerline{\psfig{figure=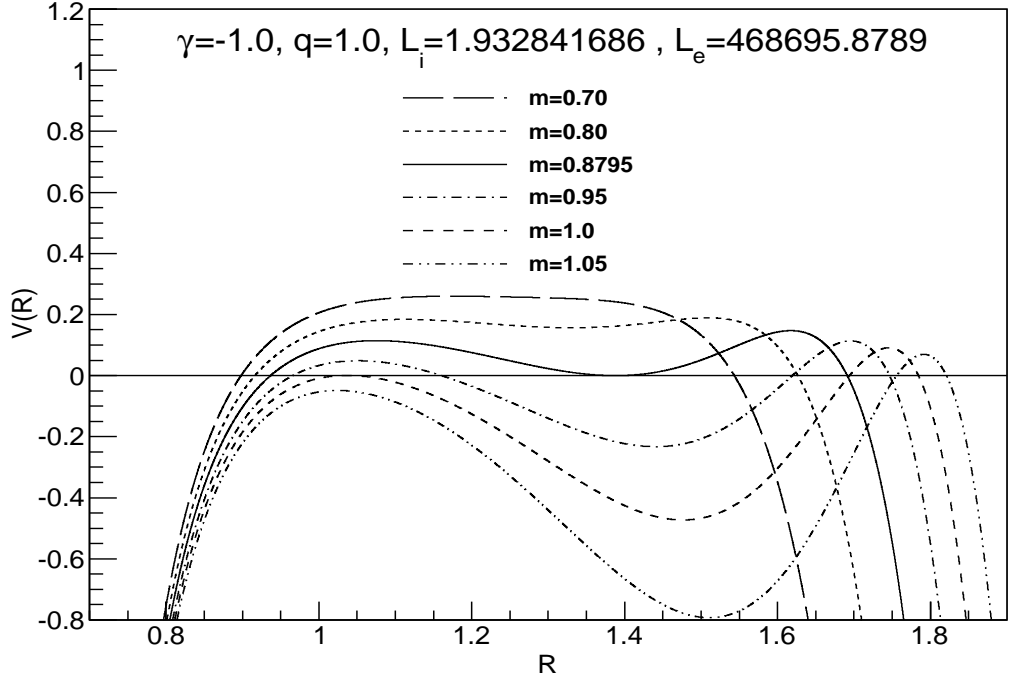,width=6.0truein,height=4.0truein}
}\caption{The potential $V(R)$ for $\gamma=-1$, $q=1$, 
$L_i=1.932841686$ and $L_e=468695.8789$. This figure shows as the mass, for a given charge, contributes for the stability of the shell.}
\label{fig1a}
\end{figure}

\begin{figure}
\vspace{.2in}
\centerline{\psfig{figure=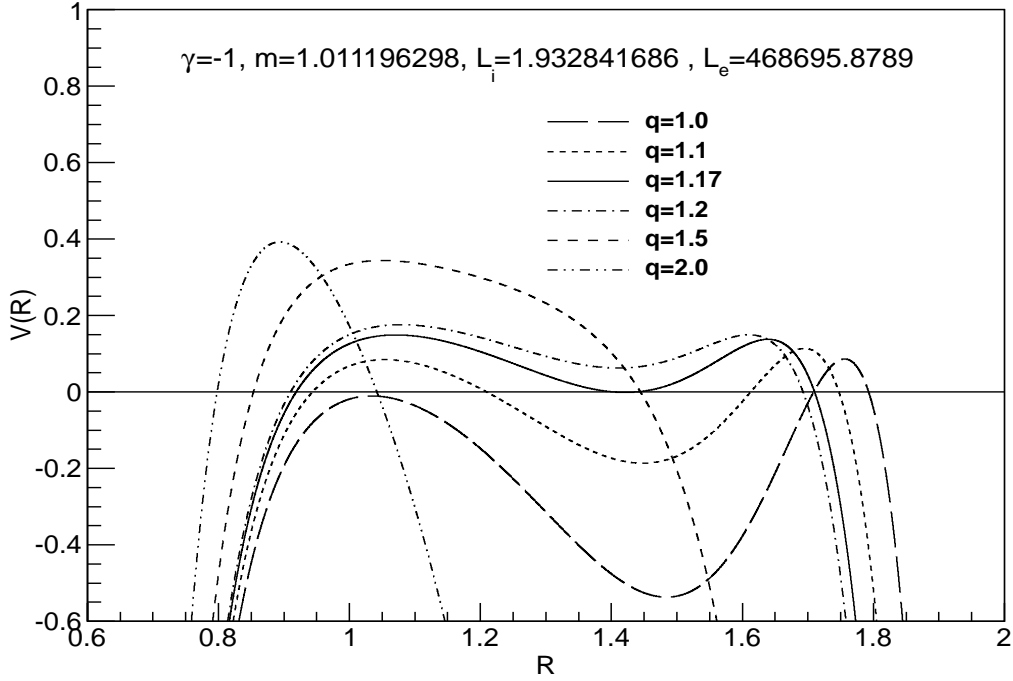,width=6.0truein,height=4.0truein}
}\caption{The potential $V(R)$ for $\gamma=-1$, $m=1.011196298$, 
$L_i=1.932841686$ and $L_e=468695.8789$. 
This figure shows the influence of the charge for the stability of the
shell. As showed in the figure 1, the increase of the charge implies in the loss of the stability.}
\label{fig1b}
\end{figure}

\begin{table}
\caption{\label{table1} This table shows the calculated horizons using
the equations (\ref{Rdsec}), (\ref{Rrnoah}) and (\ref{Rrniah}). When $q=0$ we have
used the equations (\ref{rbh}) and (\ref{rc}). Hereinafter, the symbol 
$i=\sqrt{-1}$ denotes the imaginary unit. $L_i=1.93284$, $\gamma=-1$. 
See figure \ref{fig1}.}
\begin{ruledtabular}
\begin{tabular}{cccccccc}
$m$ & $q$ & $L_e$ & $R_{rniah}$ & $R_{rnoah}$ & $R_{dsec}$ \\
    &     &       &             & $R_{bh}$    & $R_{c}$     \\
\hline
0.50559 & 0 & 468695.8789 &  & 1.01139 & 468695.3735 \\
0.50559 & 0.1 & 468695.8789 & 0.6377 & -0.6377 & 468695.8789 \\
0.50559 & 0.2 & 468695.8789 & 0.6377 & -0.6377 & 468695.8789 \\
0.50559 & 1.5 & 468695.8789 & 1.9134 & 1.9134 & 468693.9654 \\
0.50559 & 1.85323 & 468695.8789 & 1.9134 - 2.01697 $i$ & 1.9134 + 2.016916 $i$ & 468693.9654 + 0.00002 $i$ \\
0.50559 & 10 & 468695.8789 & 0.0166 - 9.89753 $i$ & 0.0166 + 9.86441 $i$ & 468695.8626 + 0.01655 $i$ \\
\end{tabular}
\end{ruledtabular}
\end{table}

\begin{figure}
\vspace{.2in}
\centerline{\psfig{figure=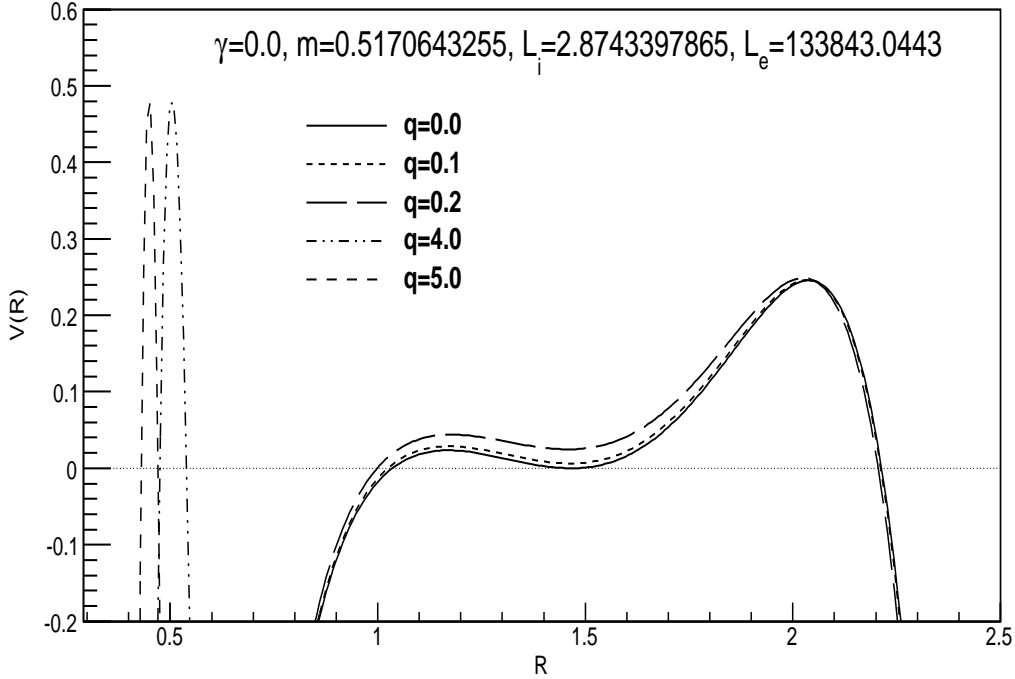,width=6.0truein,height=4.0truein}
}\caption{The potential $V(R)$ for $\gamma=0$, $m=0.5170643255$, $L_i=2.8743397865$ 
and $L_e=133843.0443$. The horizons are given in Table \ref{table2}. 
For $q=0.0$ a stable gravastar is formed if $R_0=1.5$, but for bigger charges, the shell collapses to a naked singularity.}
\label{fig2}
\end{figure}

\begin{table}
\caption{\label{table2} This table shows the calculated horizons using
the equations (\ref{Rdsec}), (\ref{Rrnoah}) and (\ref{Rrniah}). When $q=0$ we have
used the equations (\ref{rbh}) and (\ref{rc}). 
$L_i=2.87433$, $\gamma=0$. 
See figure \ref{fig2}.}
\begin{ruledtabular}
\begin{tabular}{cccccccc}
$m$ & $q$ & $L_e$ & $R_{rniah}$ & $R_{rnoah}$ & $R_{dsec}$ \\
    &     &       &             & $R_{bh}$    & $R_{c}$     \\
\hline
0.51706 & 0 & 133843.0443 &  & 1.03420 & 133842.5272 \\
0.51706 & 0.1 & 133843.0443 & 0.03125 & 1.06157 & 133842.4978 \\
0.51706 & 0.2 & 133843.0443 & 0.06453 & 1.02829 & 133842.4979 \\
0.51706 & 0.61090 & 133843.0443 & 0.77273 - 0.5759 $i$ & 0.77275 + 0.5759 $i$ & 133842.2716 + 0.000005 $i$ \\
0.51706 & 4 & 133843.0443 & 1.09282-3.94867 $i$ & 1.09282+3.9486 $i$ & 133841.9516+0.000028 $i$ \\
0.51706 & 5 & 133843.0443 & 1.09282-4.95470 $i$ & 1.09282+4.95463 $i$ & 133841.9516+0.000035 $i$ \\
0.51706 & 10 & 133843.0443 & 1.22180 - 9.97613 $i$ & 1.22182 + 9.97600 $i$ & 133841.8230 + 0.00006 $i$ \\
0.51706 & 20 & 133843.0443 & 0.01260 - 20.00625 $i$ & 0.01258 + 19.98107 $i$ & 133843.0333 + 0.01258 $i$ \\
\end{tabular}
\end{ruledtabular}
\end{table}

\begin{figure}
\vspace{.2in}
\centerline{\psfig{figure=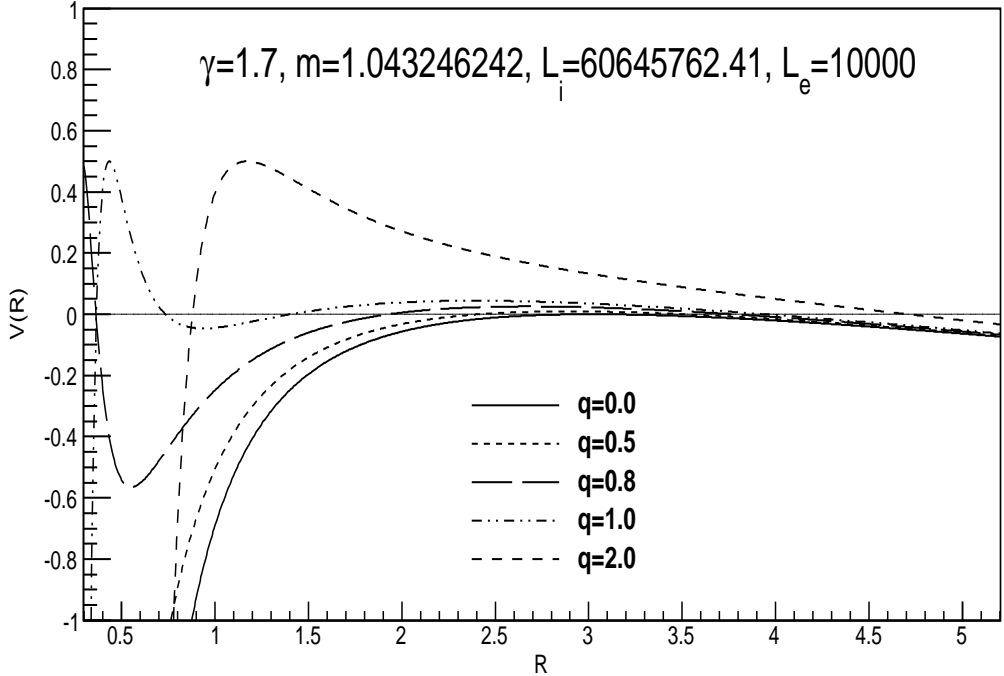,width=6.0truein,height=4.0truein}
}\caption{The potential $V(R)$ for $\gamma=1.7$, $m=1.043246242$, $L_i=60645762.41$ 
and $L_e=10000$. The horizons are given in Table \ref{table3}. 
Note that we have a bounded excursion gravastar for $q=0.8$ and $q=1.0$, and a naked singularity for $q=2.0$.}
\label{fig3}
\end{figure}

\begin{table}
\caption{\label{table3} This table shows the calculated horizons using
the equations (\ref{Rdsec}), (\ref{Rrnoah}) and (\ref{Rrniah}). When $q=0$ we have
used the equations (\ref{rbh}) and (\ref{rc}).  
$L_i=60645762.41$, $\gamma=1.7$. 
See figure \ref{fig3}.}
\begin{ruledtabular}
\begin{tabular}{cccccccc}
$m$ & $q$ & $L_e$ & $R_{rniah}$ & $R_{rnoah}$ & $R_{dsec}$ \\
    &     &       &             & $R_{bh}$    & $R_{c}$     \\
\hline
1.04324 & 0 & 10000 &  & 2.08648 & 9998.95659 \\
1.04324 & 0.5 & 10000 & 0.12762 & 1.95883 & 9998.95661 \\
1.04324 & 0.8 & 10000 & 0.37337 & 1.71308 & 9998.95664 \\
1.04324 & 1.0 & 10000 & 0.74584 & 1.33902 & 9998.95745 \\
1.04324 & 1.04385 & 10000 & 1.04403 - 0.04303 $i$ & 1.04403 + 0.04303 $i$ & 9998.95586 + 0.2440$\times 10^{-7} i$ \\
1.04324 & 2.0 & 10000 & 1.04243-1.70619 $i$ & 1.04243+1.70619 $i$ & 9998.95760 + 0.9702$\times 10^{-6} i$\\
1.04324 & 10 & 10000 & 1.04083 - 9.94541 $i$ & 1.04083 + 9.94540 $i$ & 164285.2671 + 0.00008 $i$ \\
1.04324 & 20 & 10000 & 1.04801 - 19.97270 $i$ & 1.048013 + 19.97267 $i$ & 164286.4156 + 0.01393 $i$ \\
\end{tabular}
\end{ruledtabular}
\end{table}

\begin{figure}
\vspace{.2in}
\centerline{\psfig{figure=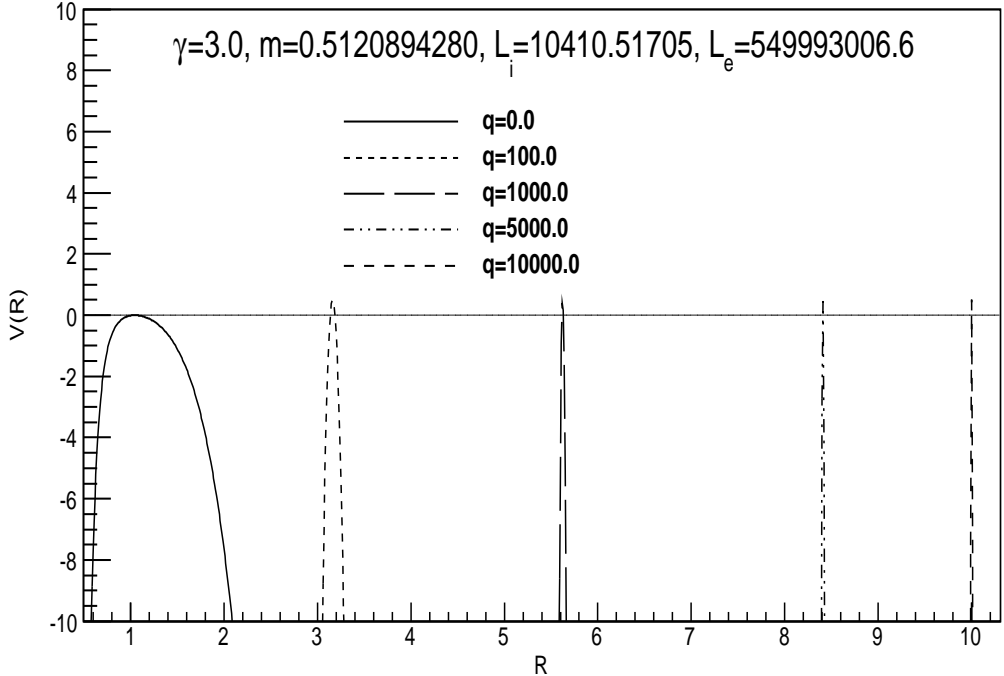,width=6.0truein,height=4.0truein}
}\caption{The potential $V(R)$ for $\gamma=3$, $m=0.5120894280$, $L_i=10410.51705$ and 
$L_e=549993006.6$. The horizons are given in Table \ref{table4}. 
In this example, the shell is constituted by phantom energy and none stable structure is formed.}
\label{fig4}
\end{figure}

\begin{table}
\caption{\label{table4} This table shows the calculated horizons using
the equations (\ref{Rdsec}), (\ref{Rrnoah}) and (\ref{Rrniah}). When $q=0$ we have
used the equations (\ref{rbh}) and (\ref{rc}).  
$L_i=10410.51705$, $\gamma=3$. 
See figure \ref{fig4}.}
\begin{ruledtabular}
\begin{tabular}{cccccccc}
$m$ & $q$ & $L_e$ & $R_{rniah}$ & $R_{rnoah}$ & $R_{dsec}$ \\
    &     &       &             & $R_{bh}$    & $R_{c}$     \\
\hline
0.51208 & 0 & 549993006 &  & 1.06648 & 54993005.47 \\
0.51208 & 100 & 549993006 & 0.51353 - 99.99419 $i$ & 0.51353+99.99395 $i$ & 10410.48370+0.00011 $i$\\
0.51208 & 1000 & 549993006 & 0.50823-995.46055 $i$ & 0.50823+995.45813 $i$ & 10457.4936 + 0.00121 $i$ \\
0.51208 & 2049.7030 & 549993006 & 2245.2 - 2366.8273 $i$ & 2245.4 + 2366.7587 $i$ & 0.5499$\times 10^{9} + 0.03432 i$ \\
0.51208 & 5000 & 549993006 & 0.38728-4577.13222 $i$ & 11371.90700-0.00795 $i$ & 0.38728+4577.14813 $i$ \\
0.51208 & 10000 & 549993006 & 0.25613-7948.23843 $i$ & 13097.59644-0.02406 $i$ & 0.25613+7948.28656 $i$ \\
\end{tabular}
\end{ruledtabular}
\end{table}

\begin{figure}
\vspace{.2in}
\centerline{\psfig{figure=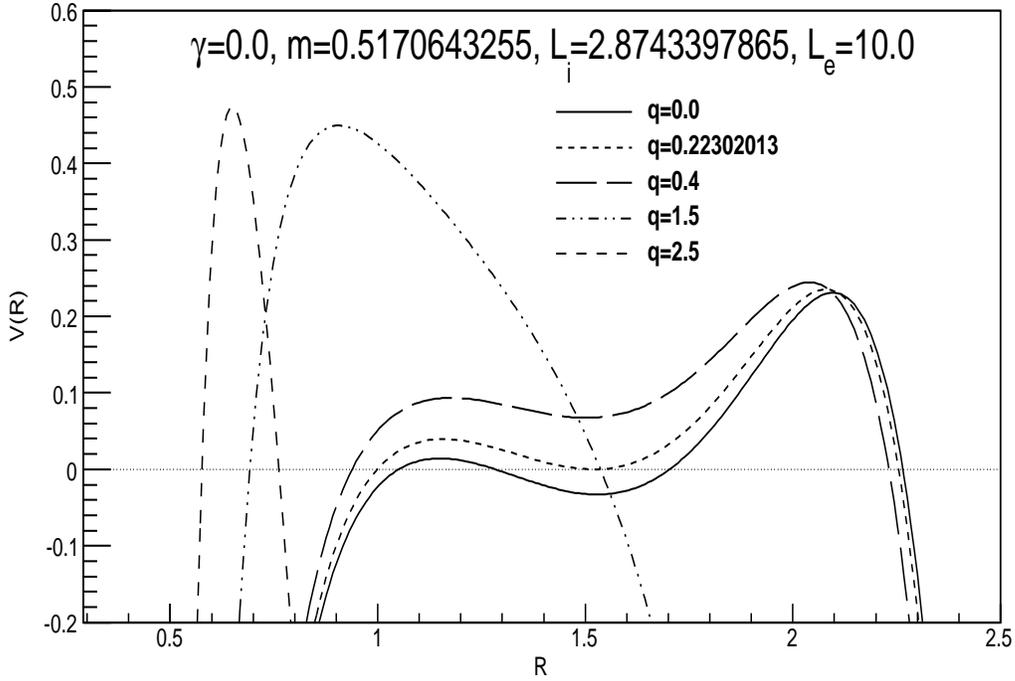,width=6.0truein,height=4.0truein}
}\caption{The potential $V(R)$ for $\gamma=0$, $m=0.5170643255$, $L_i=2.8743397865$ and 
$L_e=10.0$. The horizons are given in Table \ref{table5}. 
Note that we have a gravastar enclosing a naked singularity.}
\label{fig5}
\end{figure}

\begin{table}
\caption{\label{table5} This table shows the calculated horizons using
the equations (\ref{Rdsec}), (\ref{Rrnoah}) and (\ref{Rrniah}). When $q=0$ we have
used the equations (\ref{rbh}) and (\ref{rc}).  
$L_i=2.8743397865$, $\gamma=0$. 
See figure \ref{fig5}.}
\begin{ruledtabular}
\begin{tabular}{cccccccc}
$m$ & $q$ & $L_e$ & $R_{rniah}$ & $R_{rnoah}$ & $R_{dsec}$ \\
    &     &       &             & $R_{bh}$    & $R_{c}$     \\
\hline
0.51706 & 0 & 10 &  & 1.04554 & 9.43614 \\
0.51706 & 0.22303013 & 10 & 0.05056 & 0.99390 & 9.43929 \\
0.51706 & 0.4 & 10 & 0.18939 & 0.85268 & 9.44625 \\
0.51706 & 1.5 & 10 & 0.49983 - 1.40462 $i$ & 0.49983 + 1.40462 $i$ & 9.573556+ 0.14 $\times 10^{-8}$ $i$ \\
0.51706 & 2.5 & 10 & 0.77273 - 0.57597 $i$ & 0.77275 + 0.57596 $i$ & 133842.2716 + 0.59067 $\times 10^{-5}$ $i$ \\
\end{tabular}
\end{ruledtabular}
\end{table}

\section{Conclusions}
We constructed a gravastar model  consisting by an interior de Sitter spacetime and an exterior
spacetime with an external cosmological constant, described by a de Sitter-Reissner-Nordstr\"om metric. 
The charge is localized on the shell. Restricting the range of the initial radius, we obtain as the final structure bounded excursions,
stable gravastars and also naked singularities. 

We investigated the influence of the charge and we observed that increasing its value, and fixing a value for the mass,
we can obtain a stable gravastar from a bounded excursion gravastar.  For even higher values of the charge
the apparent horizon can be avoided, which leads the formation of a naked singularity.
  
In the case of a shell formed by standard energy, with $\gamma = -1$ in its state equation (figures 1, 2 and 3), for some fixed values of charge, 
above a lower limit $q^*$ dependent on the mass and on cosmological constants, and small initial values of radius, the shell  can collapse and forms a naked singularity. There is also a possibly formation of  black holes for some values of the charge, 
below $q^*$, for fixed values of the mass. Increasing the value of the charge, we verify that initially bounded 
excursion configurations become more stable and there is another limit for the charge, $q_c$, for  which the structure becomes
a stable gravastar. For charges upper than $q_c$ the shell collapses. Moreover,  fixing the mass and varying the charge,  we have a similar 
behavior, that is, a bounded excursion becomes more stable increasing the charge until reaches a stable gravastar and for others 
values of the charge, we have black holes and naked singularities. In the case of $\gamma=1.7$ (figure 5), the shell is constituted by dark 
energy and for small values of the initial radius there is also a naked singularity formation.
The same is found for a phantom dark energy shell with $\gamma = 3.0$ (figure 6). Finally, for a stiff fluid shell with $\gamma = 0 $ (figures 4 and 7), we have 
bounded excursion formation, stable gravastar, black hole and naked singularity formation according to the values of the charge and for 
some values of initial radius.
It is remarkable that the naked singularity formation appeared in this gravastar model is completely new. Then, beginning with a shell linking two spacetimes (de Sitter and de Sitter-Reissner-Nordstr\"om) in order to eliminate the horizons, 
as proposed by the gravastar's model, except for the cosmological horizon of the exterior spacetime, the shell can stay stable, 
forming a gravastar, or collapsing in a black hole or even a naked singularity, representing a new counterexample to the Cosmic Censorship.
Then, this model definitively is not an alternative to the black hole,  even naked singularities.

\begin{acknowledgments}
The financial assistance from FAPERJ/UERJ (CFCB and MFAdaS) are gratefully 
acknowledged.
The authors (RC, MFAdaS) acknowledge the financial support from FAPERJ 
(no. E-26/171.754/2000, E-26/171.533/2002, E-26/170.951/2006, E-26/110.432/2009 
and E-26/111.714/2010). The authors (RC and MFAdaS) also acknowledge the 
financial support from Conselho Nacional de Desenvolvimento Cient\'ifico e
Tecnol\'ogico - CNPq - Brazil (no. 450572/2009-9, 301973/2009-1 and 
477268/2010-2). The author (MFAdaS) also acknowledges the financial support 
from Financiadora de Estudos e Projetos - FINEP - Brazil (Ref. 2399/03). 
\end{acknowledgments}

\end{document}